\documentclass[12pt,preprint]{aastex}

\usepackage{natbib}
\begin{document}

\title{On the size, shape, and density of dwarf planet Makemake}
\author{M.E. Brown}
\affil{Division of Geological and Planetary Sciences, California Institute
of Technology, Pasadena, CA 91125}
\email{mbrown@caltech.edu}

\begin{abstract}
A recent stellar occultation by dwarf planet Makemake
provided an excellent opportunity
to measure the size and shape of one of the largest
objects in the Kuiper belt. The analysis of these results
provided what were reported to be precise measurements of the
lengths of the projected axes, the albedo, and even the density
of Makemake, but these results were, in part, derived from
qualitative arguments.
We reanalyzed the occultation timing data
using a quantitative
statistical description, and, in general,
find the previously reported results on the shape of Makemake
to be unjustified.
In our solution, in which we use our inference
from photometric data that Makemake is being viewed
nearly pole-on, we find a 1$\sigma$ upper limit to the projected
elongation of Makemake of 1.02, with measured equatorial diameter
of 1434$\pm$14 km and a projected polar diameter of 1422$\pm$14 km, yielding
an albedo of 0.81$^{+0.01}_{-0.02}$. If we remove the external
constraint on the pole position of Makemake, we find instead a 1$\sigma$
upper limit to the elongation of 1.06, with a measured equatorial diameter of 
1434$^{+48}_{-18}$ km and a projected polar diameter of 1420$^{+18}_{-24}$ km,
yielding an albedo of 0.81$^{+0.03}_{-0.05}$.
Critically, we find that
the reported measurement of the density of Makemake was
based on the misapplication of the volatile retention models.
A corrected analysis shows that the
occultation measurements provide no meaningful constraint on the density
of Makemake.
\end{abstract}

\section{Introduction}

The density of a solar system body is one of the most important
parameters for understanding the composition, evolution, and formation
history of the object. In the Kuiper belt, the wide range of
densities, from below that of water ice to that of nearly pure
rock, is one of the mysteries that continues to have no
satisfactory explanation. \citet{2012AREPS..40..467B} proposed several
classes of general solutions to explain, in particular, the
wide range of densities of objects in the dwarf planet size
range. In one limiting scenario, densities gradual increase
with size as small amounts of ice are removed with each
accretional impact. In the other limit, the densities of
the largest objects are stochastically set by single giant
impacts which can remove significant quantities of ice and
lead to one or more satellites with a small fraction of the mass
of the primary.
The density of Makemake could be a key discriminator between
these types of models. Makemake is the largest known Kuiper belt
object for which no satellite has been detected 
\citep{2006ApJ...639L..43B, 
2008ssbn.book..335B}.
In that case, Makemake might never have suffered a density-increasing
giant impact, and thus could have a density lower than the typical
values of $\sim$2.1 g cm$^{-3}$ and higher that appear typical for dwarf
planets with small satellites. In contrast, a density $\gtrsim 2.1$ 
g cm$^{-3}$ for Makemake would indicate the correlation between
high densities and the presence of collisional satellites is a mere
coincidence unrelated to formation.

\citet[][hereafter O12]{2012Natur.491..566O} measured a stellar occultation of
Makemake and, from these data, infer a density of 1.7$\pm$0.3
g cm$^{-3}$ for the object. 
Such a density would strongly support the classes of
models in which the high densities of objects like 
Haumea and Eris are due to single giant impacts which left
small moons in orbit.
Because of the importance of
this density for constraining the formation pathways of
these icy dwarf planets, we investigate this density measurement
to determine its robustness.
To do so we reanalyze the occultation data of O12 using a quantitative,
rather than qualitative,
statistical framework. In addition to examining the density,
this new analysis allows us to obtain
statistically justifiable constraints
on the size, shape, and albedo of Makemake for the first time.

\section{Observations and spherical fitting}
Makemake occulted the star NOMAD 1181-0235723 on 23 April 2011. 
O12 report 8 detections of the occultation
from stations in Chile and Brazil, and they fit square-well occultation
models to determine the time of stellar disappearance and reappearance
for each station, as well as uncertainties. The data quality are
exquisite, with event uncertainties as small as 0.04 seconds in the
best case, corresponding to chord lengths with uncertainities of only
a few kilometers. As seen in Fig. 1, however, one difficulty
with the data is that 5 of the 8 chords sample nearly the same
region on Makemake and the three remaining stations
sample identical chords 300 km south (Fig. 1). The lack
of strong constraint on the north-south dimension 
dominates the shape results of O12.

To determine the shape and density of Makemake,
O12 first fit the data using simple $\chi^2$ minimization
and then present a series of qualitative
arguments based on additional consideration to modify the results
given by the data. Such an approach need not remain qualitative but
can be given statistical meaning by adopting a Bayesian approach.
We develop such an approach here.

To check for consistency with O12, we first attempt to reproduce
their basic results and determine the
best-fit model to describe the data assuming that Makemake is perfectly
spherical. In this case we fit three 
parameters: the sphere diameter, $d$, and the $x$ and $y$ offset 
of the shadow of Makemake from 
the center of the predicted path. In our analysis we calculate
predicted disappearance and appearance times at each station given
values of $d$, $x$, and $y$, and we compute the likelihood,
which is identical to the value of $\chi^2$ as described in O12, 
for those parameters.
To compute the Bayesian probability function, we multiply the
likelihood by the priors on all of the parameters. To begin,
we assume simple uniform
priors in $x$, $y$, and $d$ for best comparison to O12.

To determine the probability distribution function (PDF) for each of
the parameters, 
we integrate through phase space
using a 
Markhov Chain Monte Carlo (MCMC) scheme. We use the Python 
package {\it emcee} \citep{2012arXiv1202.3665F} which implements the 
\citet{ISI:000282653600004}  affine invariant ensemble sampler for MCMC.
For our simple spherical fit, we found good convergence using an ensemble
of 100 chains running 10$^4$ steps with a initialization
(``burn-in'') period, which is discarded,
 of 10\% of the total length of each chain.
The $x$ and $y$ offsets of the star are of no interest, so
we treat them as nuisance parameters and marginalize over 
their distributions. The distribution of $d$ -- that is, the PDF 
marginalized over the other two parameters -- is nearly Gaussian,
and we find that the
spherical diameter is $1430\pm7$ km. 
Throughout this paper we define the best fit as the peak of the 
PDF and the 1$\sigma$ range as the smallest region about the
best fit containing 68.2\% of the probability. If the peak of the PDF 
is at or near one of the extrema we report an upper or lower limit 
with the same method.
The modest improvement
in the uncertainty from the O12 result of $1430\pm9$ km is the result
of the marginalization and is a small demonstration of the usefulness
of this technique.  
\begin{figure}
\plotone{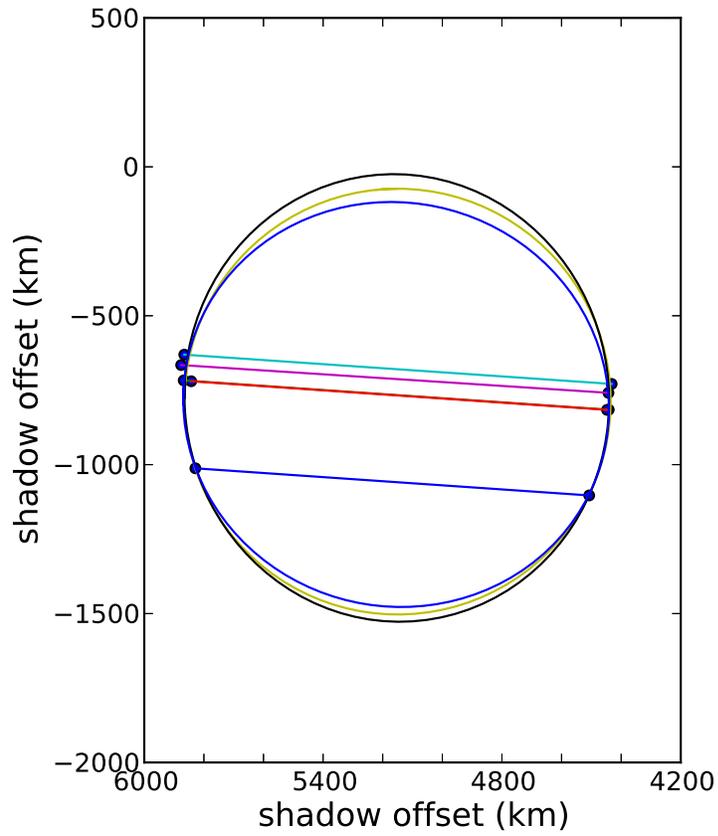}
\caption{The projected locations of the stellar disappearance and reappearance
at each of the observing stations. The lack of stations further north and
south lead to weak constraints on the elongation in that direction. 
We show the best fit circular shadow, with a diameter of $1430\pm7$ km,
as well as shadows with the 1$\sigma$ maximum elongations allowed
when using the density prior. }
\end{figure}

\section{Ellipsoid fit}
As correctly pointed out by O12, given plausible densities and
the measured 7.77 hour spin
period of Makemake \citep{2009AJ....138..428H}, the true shape of Makemake will
not be a sphere, but rather a Maclaurin spheroid with an
elongation dependent on the density and spin rate
\citep{1969efe..book.....C}. Such a spheroid,
viewed in projection, will appear as an ellipse. 
O12 fit
directly to an elliptical shape and find, not surprisingly, that
their
best-fit ellipse is elongated in the north-south direction, with
an axial ratio of 1.15$\pm$0.17. They then argue
that a true elongation in the north-south direction would be 
coincidental and that the real elongation is probably
smaller. Qualitative arguments are given to suggest a ``preferred''
elongation of 1.05$\pm$0.03, though the reasons for 
these precise values are unclear.

These arguments can be approached
statistically rather than qualitatively. Rather than fit an
ellipse to the projected shape of the body, we fit
to the full Maclaurin spheroid shape. We parameterize this shape
with four parameters, $E$, 
the ratio of the equatorial
diameter to the polar diameter, $d$, the polar diameter, $\phi$, the
angle of the pole with respect to the line-of-sight (which we call the
``polar axis angle''), and, $\theta$, 
the position angle of the largest dimension of
the projected ellipse (the ``azimuthal angle'').
Such a fit has many degeneracies; these degeneracies, rather than
being a problem, correctly account for the volume of phase space in
the multi-dimensional fit and give a correct accounting of 
the probabilities of each of these parameters.

In our ellipsoid fit, we constrain
the ratio of the polar to equatorial diameters  to
be between 1 and 1.716, the minimum and maximum values
obtainable by a Maclaurin spheroid \citep{1969efe..book.....C} (we add 
additional constraints on the shape below). We add no constraints on
the azimuth angle, and we chose $\phi$ such that
the polar axis is oriented arbitrarily in space (we also modify
this constraint below). 
Because of the much larger phase space to be explored, we run our
MCMC sampler with an ensemble of 100 chains sampled through
$10^5$ iterations.

We find that the PDF
of the elongation, $E$, is not highly constrained. The PDF
for $E$
peaks at $E=1.03$
and decreases to 20\% of the peak value by 1.716.
The 1$\sigma$ upper limit on the true elongation
is 1.37.
The projected elongation, 
is, however, more tightly constrained.
The PDF peaks at 1.0 and has a 1$\sigma$ upper
limit of of only 1.09 (Figure 2).
Simply by correctly using
our prior knowledge that we are looking at a two-dimensional
projection of an arbitrarily oriented Maclaurin spheroid, 
this analysis provides a three times
tighter constraint on the measured projected elongation than
the O12 1$\sigma$ upper limit of 1.32. 
Our PDF for $\phi$ is not 
 as would be expected for a random orientation,
but rather has a peak and 1$\sigma$ range of $20^{+40}_{-8}$.
\begin{figure}
\plotone{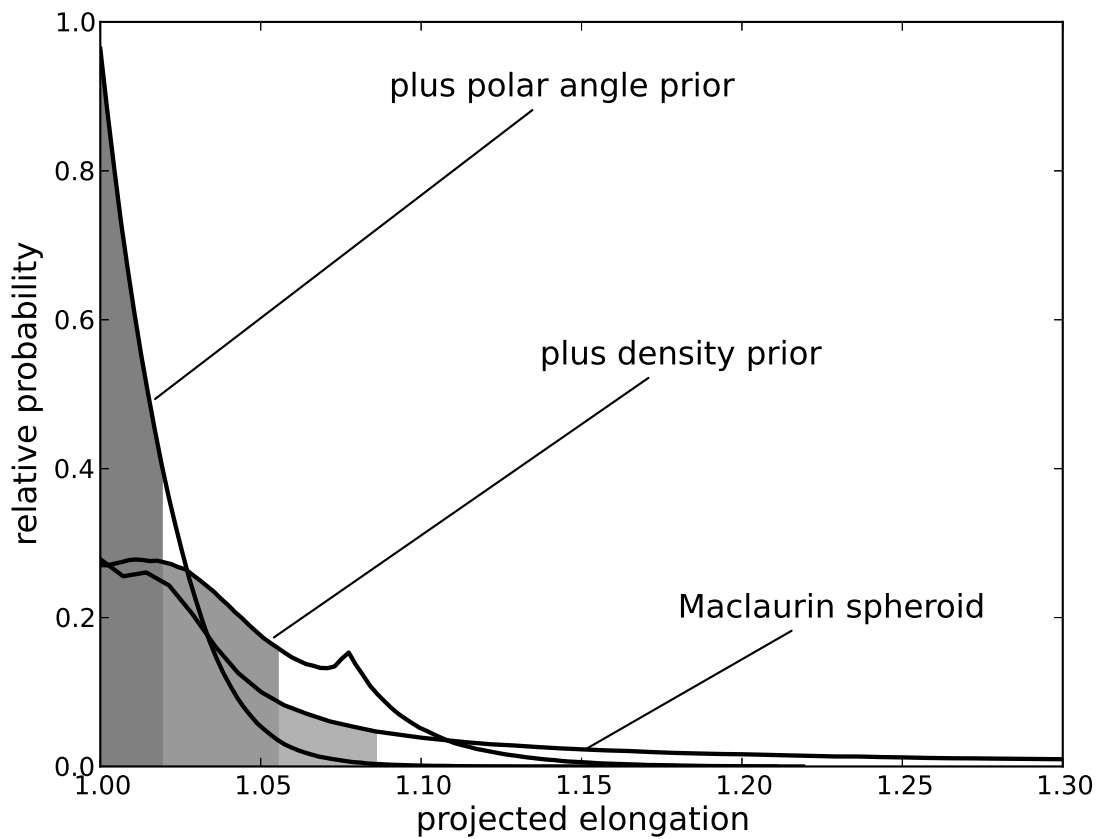}
\caption{The PDF of the projected elongation.
The grey shaded regions show the 1$\sigma$ regions containing 68.2\%
of the probability density. While the Maclaurin spheroid constraint
yields a long
tail to large projected elongations, adding the density
prior with a  lower limit of
1.3 g cm$^{-3}$ 
limits the maximum true elongation
and thus the projected elongation. A prior on the polar angle
assuming a nearly pole-on view of Makemake yields even smaller
projected elongations.}
\end{figure}

These results
correctly take into
account both the constraints from the data and the geometric
expectations of a randomly oriented projected ellipse. 
Knowing that the Maclaurin spheroid can
have a wide range of values for $E$ between 1 and 1.716, 
the pole position must be close
to the line-of-sight or else a larger projected elongation would be measured,
 {\it or} the elongation must be close to the north-south direction
where the elongation is unconstrained.
But, because the volume of parameter space is low in the unconstrained
direction, the overall likelihood of this orientation and thus
these large elongations is small.
Figure 3 illustrates this effect by showing probability contours
of elongation azimuth vs. projected elongation. 
\begin{figure}
\plotone{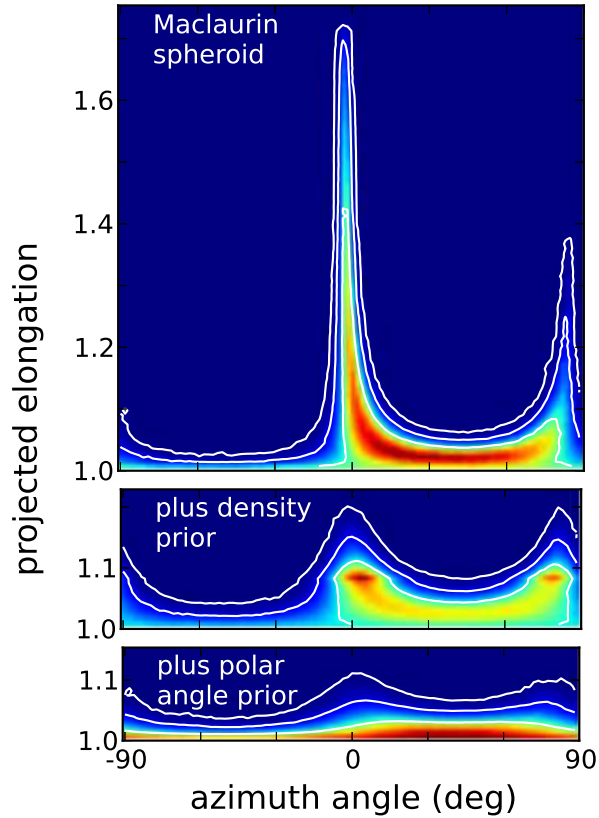}
\caption{Two-dimensional probability densities of azimuth angle
vs. projected elongation (the projected elongation PDFs of
Fig. 1 are integrals of these two dimensional functions over
azimuth angle). In the Maclaurin spheroid fit and the fit with
the density prior added, the highest probability density 
regions have either low elongation or have a elongation
aligned in the unconstrained north-south direction. Note that both
angles near 0 degrees -- with their major axis aligned north-south --
angles near $\pm$90 degrees -- with their minor axis aligned
north-south -- result from this lack strong north-south constraint.
Both configurations yield acceptable fits (see Fig. 1).
If the polar angle is assumed to be small, the preference for north-south
elongation nearly disappears as the projected elongation is small.}
\end{figure}

\section{Fit to radiometric and density constraints}
We now add prior constraints
to help us determine the
projected shape. 
We first use the combined Spitzer and Herschel
radiometry which suggests that the projected area of Makemake is
1420$\pm$60 km \citep{2010A&A...518L.148L}. 
This prior mainly serves to limit very large elongations
in the north-south direction which would cause the thermal emission
to be higher than observed.

A more important prior is on the density. While above,
we allowed $E$ to range between 1 and 1.716 --
essentially choosing a uniform prior for the elongation -- a more
physically motivated approach is to use a prior on the density
itself, rather than the elongation, and allow
the density and rotation period to determine the elongation.

O12 claim that their occultation results
provide a measurement of the density of Makemake of 1.7$\pm$0.3
g cm$^{-3}$. 
The primary justification for the assumption of
a density of 1.7$\pm$0.3 g cm$^{-3}$ is that for higher densities,
the volatile retention models of 
\citet{2007ApJ...659L..61S} \citep[updated in]{2011ApJ...738L..26B} 
predict retention of N$_2$ for a body of the size and temperature
of Makemake, which
would then result in a measurable atmosphere in the O12 occultation
data. This statement is,
however, a gross misinterpretation of the volatile retention
models. In the models of \citet{2007ApJ...659L..61S}, we explicitly and
deliberately calculate the {\it slowest} possible volatile loss
mechanism -- Jeans escape -- so that objects that would
lose all of their volatiles due to this mechanism
must have lost all of their accessible volatiles. But
it is incorrect to say that objects which could have retained 
volatiles against Jeans escape must have retained them against
all other mechanisms. Haumea is an excellent example. 
Based on its size, mass, and temperature, it could easily
hold its volatiles against Jeans escape over the life of the 
solar system. Another process, however, 
presumably a giant impact \citep{2007Natur.446..294B}, 
led to complete volatile loss. The arguments of O12 would
instead state that the lack of CH$_4$ on Haumea constrains
its density to be $\sim$1 g cm$^{-3}$, rather than its measured
value of $\sim$2.6 g cm$^{-3}$ 
\citep{2006ApJ...639.1238R}!
These arguments for the density constraint on Makemake
based on the absence of a detectable atmosphere
are clearly spurious and should be given no weight.

The O12 lower limit to the density of 1.4 g cm$^{-3}$ comes
from using the same volatile retention models to explain the
continued presence of CH$_4$ on the surface. Here the volatile loss
model is used correctly. For densities below 1.4 g cm$^{-3}$ 
Makemake must have lost all of its CH$_4$ even if the only
escape process was slow Jeans escape. We retain this
lower limit as a sensible constraint.

The upper limit to the density from O12 comes from assuming that the
inferred value of 1.7 g cm$^{-3}$ is close to correct and positing that
objects of similar size have densities within about 0.3 g cm$^{-3}$
of that value. 
Use of this upper limit
is problematic. 
The most important question to address about the density of
Makemake is whether it has a value below about 2.1 g cm$^{-3}$, as
might be typical for objects with large, potentially captured
satellites, or a higher value as
might be typical for objects with evidence for giant impacts.
Assuming that the upper limit to the density of Makemake is
2.0 g cm$^{-3}$ simply asserts an answer to that question,
which is clearly unacceptable.

Based on these considerations, 
we retain the 1.4 g cm$^{-3}$ lower
limit based on the retention of CH$_4$ (though we 
employ a softer cutoff by assuming a one-sided Gaussian
distribution with a $\sigma$ of 0.03 for densities below
1.4 g cm$^{-3}$), and we add the only reasonable upper limit
density that we can find, which is that the density is certainly
-- we assume -- below that of solid rock, or about 3.2 g cm$^{-3}$. 
We see no justification for placing any other prior constraints
on the density of Makemake, so from 1.4 to 3.2 g cm$^{-3}$ we
assume a uniform prior. 

Once again employing our MCMC integration, 
we find that when using the density prior the PDF for
the projected elongation peaks strongly at 1.0, with a 1$\sigma$ upper
limit of 1.06 (Fig. 2). The smaller elongation derived here is the
result of the 1.3 g cm$^{-3}$ lower limit on the density, 
which, for a 7.77 hour
rotation period, translates to
an upper limit to the true elongation, $E$, of 1.20. For the same reason,
the polar axis angle is not as strongly constrained to
small values, and we find a best fit at $32^{+23}_{-19}$.
The projected polar diameter has values of $1420^{+18}_{-24}$ km.
By calculating the PDF of the projected area and comparing it
to the results of O12, we also calculate an albedo PDF
with a distribution peak and range of $0.81^{+0.03}_{-0.05}$.

The PDF for the density 
rises linearly from
1.4 to 3.2 g cm$^{-3}$ (Fig. 4). 
We can derive a formal 1$\sigma$ lower limit
to the density of 2.14 g cm$^{-3}$, but this
lower limit is meaningless, as the distribution 
of the density is just the density prior
modified to have a slight preference for less elongation
and thus higher densities. 
The data themselves
provide nearly no constraint on the density. 
\begin{figure}
\plotone{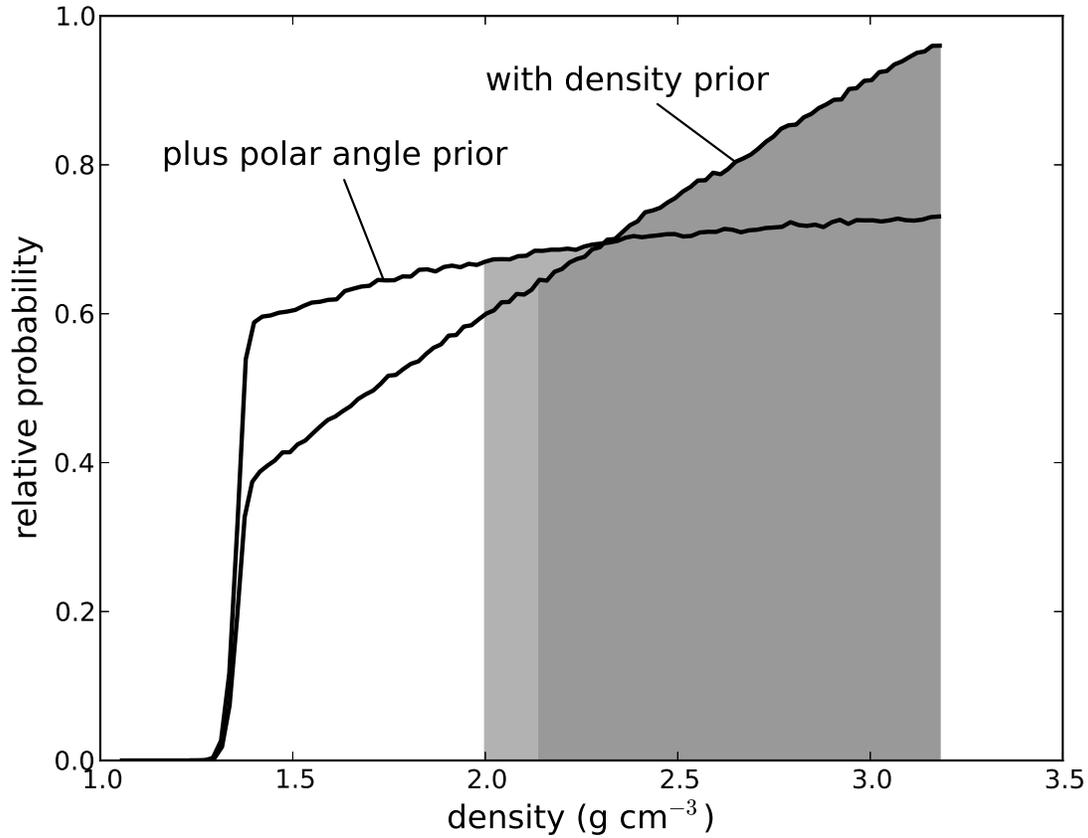}
\caption{The PDF for density. The prior on 
density is uniform from 1.3 to 3.2 g cm$^{-3}$.
The occultation data suggest that higher 
densities are preferred owing to their smaller true
elongations, but, as the PDF shows, the constraint is
nearly meaningless. Adding the prior that Makemake
is viewed nearly pole-on puts less constraint on the 
elongation and thus on the density. The density PDF very
nearly resembles the original prior in this case, showing
that, again, the density is unconstrained by the occultation
data.}

\end{figure}

\section{Polar axis constraints}
As discussed by O12, there is reason to believe that we
are viewing Makemake nearly pole-on. Spitzer radiometry can
only be fit by assuming that the surface of Makemake -- like that
of Pluto --
contains a combination of very high and very low albedo 
regions \citep{2008ssbn.book..161S,2010A&A...518L.148L}, yet Makemake shows 
only a 0.03 magnitude
 variation over its
7.77 hour rotation period \citep{2009AJ....138..428H}. 
Either the dark regions of Makemake
must be extremely symmetric with respect to the pole, which is
not the case for the similarly mottled Pluto \citep{2010AJ....139.1128B}
and requires special pleading, or we
are viewing Makemake from a nearly pole-on position. 

Adding a prior on the pole position of Makemake strongly
affects the results, yet there is no 
obvious statistical distribution that one should
adopt. We make a judgement based on the evidence above
that the polar axis angle is likely less than 20 degrees. We
quantify this as a prior on polar axis angle of the form
of a Gaussian distribution peaked at zero degrees, with a 
$\sigma$ of 20 degrees. This prior is in addition to the 
prior that the axes are otherwise randomly oriented in space.
The specific values cannot be statistically
justified, but will nonetheless serve as an instructive example
of how to incorporate our expectations of the polar angle. 

We once again construct the 6-dimensional PDF from an MCMC analysis.
We find, not surprisingly, that the
polar axis angle is much more tightly confined to small
angles, with the PDF peak and range of 19$\pm$11.
The projected polar radius has a value of
1422$\pm$14km, while the albedo is constrained to $0.80^{+0.02}_{-0.01}$.
The density remains nearly unconstrained, with a 
PDF that again rises linearly from 1.4 to 3.2 g cm$^{-3}$ (Fig. 4). With
our prior expectation of a small polar angle, there is less of
a bias towards high densities, as small polar angle give small projected
elongations regardless of the real elongation. In this case the
formal 1$\sigma$ lower limit on the density is 1.98 g cm$^{-3}$, but,
as before, the constraint is dominated by the prior and the data themselves
give little information on the density.
Table 1 summarizes the results of these analyses.

\section{Conclusions}
We have developed a new statistically rigorous approach to
the analysis of dwarf planet occultation data which 
incorporates our knowledge of shapes of equilibrium,
spherical geometry, and volatile retention
to place statistically justifiable limits on the shapes
of these objects. We apply the new technique to the occultation
of Makemake observed by \citep{2012Natur.491..566O}, which
was initially analyzed using a combination of quantitative
and qualitative approaches.
We find that the ``preferred'' solution of \citep{2012Natur.491..566O} 
for the elongation of Makemake is
unjustified and leads to incorrect estimates of the precise
dimensions and albedo of Makemake.  
In our solution  in which we use the inference
from photometric data that Makemake is being viewed
nearly pole-on, we find a 1$\sigma$ upper limit to the projected
elongation of Makemake of 1.02, with measured equatorial diameter
of 1434$\pm$14 km and a projected polar diameter of 1422$\pm$14 km, yielding
an albedo of 0.81$^{+0.01}_{-0.02}$. If we remove the external
constraint on the pole position of Makemake, we find instead a 1$\sigma$
upper limit to the elongation of 1.06, with a measured equatorial diameter of 
1434$^{+48}_{-18}$ km and a projected polar diameter of 1420$^{+18}_{-24}$ km,
yielding an albedo of 0.81$^{+0.03}_{-0.05}$.

The uncertainties reported here (and by O12)
are purely statistical. True shape uncertainties
are also affected by deviations from our modeled equilibrium
shapes. Our best knowledge of the shapes of icy bodies
of this diameter comes from the medium-sized icy moons of Saturn.
With the exception of Iapetus, for which a complex thermal and
rotational history within the Saturnian system has been evoked, 
deviations from equilibrium shapes of like-sized Saturnian 
satellites are of the order of 1 km \citep{Thomas2007573}. Such shape
are well below our statistical errors 
and so do not strong affect the reported results.

The density measurement of \citep{2012Natur.491..566O} is based 
on misapplication
of the volatile retention models of \citep{2007ApJ...659L..61S} and 
arbitrary assumptions about plausible densities. Unfortunately,
while the occultation measurements of O12, when analyzed correctly,
provide excellent constraints on the size, shape, and albedo
of Makemake, they contain essentially no information about density
of this object. The density of Makemake, while an 
important parameter for understanding the evolution of
the population of dwarf planets, remains unknown.

\clearpage

\begin{deluxetable}{lccccccc}

\tablecaption{Table 1}
\tablenum{1}
\tablehead{\colhead{parameter} & \colhead{sphere} & \colhead{Maclaurin} & \colhead{plus density} & \colhead{plus polar} & \colhead{} & \colhead{} & \colhead{} \\ 
\colhead{} & \colhead{} & \colhead{spheroid} & \colhead{prior} & \colhead{angle prior} & \colhead{} & \colhead{} & \colhead{} \\ }

\startdata
projected elong&1.00&     	$< 1.09$&		$< 1.06$&		$< 1.02$\\
actual elong, $E$&	1.00&    	$< 1.37$&		$1.077<E<1.12$&	$1.077<E<1.13$\\
equatorial diameter (km)&		$1430\pm14$&	1432$^{+84}_{-24}$&	1434$^{+48}_{-18}$&	1434$\pm$14\\
projected polar diameter, $d$ (km)&n/a&     	1424$^{+16}_{-14}$&	1420$^{+18}_{-24}$&	1422$\pm$14\\
actual polar diameter (km)&	n/a&     	1420$^{+20}_{-240}$&	1320$^{+40}_{-60}$&	1320$^{+20}_{-60}$\\
albedo&		0.80$\pm0.01$&      	0.80$^{+0.05}_{-0.07}$&	0.81$^{+0.03}_{-0.05}$&	0.81$^{+0.01}_{-0.02}$\\
polar angle, $\phi$ (deg)&	n/a&      	20$^{+40}_{-8}$&	32$^{+23}_{-19}$&	19$\pm$11\tablenotemark{*}\\
density&	n/a&     	n/a&			$> 2.14$\tablenotemark{*}&		$>1.98$\tablenotemark{*}\\
\enddata
\tablenotetext{*}{result dominated by prior}
\end{deluxetable}

\end{document}